\documentclass[aps,prl,twocolumn,showpacs,preprintnumbers,superscriptaddress]{revtex4}
\usepackage{amssymb}
\usepackage{graphicx}
\usepackage{dcolumn}
\usepackage{bm}

\input{tcilatex}

\begin{document}

\title{Local Distinguishability and Schmidt Number of Orthogonal States}
\author{Ping-Xing Chen}
\affiliation{Key Laboratory of Quantum Communication and Quantum Computation, University
of Science and Technology of China, Hefei 230026, People's Republic of China}
\affiliation{Department of Physics, National University of Defense Technology, Changsha,
410073, People's Republic of China}
\author{Wei Jiang}
\affiliation{Key Laboratory of Quantum Communication and Quantum Computation, University
of Science and Technology of China, Hefei 230026, People's Republic of China}
\author{Zheng-Wei Zhou}
\affiliation{Key Laboratory of Quantum Communication and Quantum Computation, University
of Science and Technology of China, Hefei 230026, People's Republic of China}
\author{Guang-Can Guo}
\affiliation{Key Laboratory of Quantum Communication and Quantum Computation, University
of Science and Technology of China, Hefei 230026, People's Republic of China}
\date{\today}

\begin{abstract}
Now, the known ensembles of orthogonal states which are distinguishable by
local operators and classical communication (LOCC) satisfy the condition
that the sum of Schmidit numbers of the orthogonal states is not bigger than
the dimensions of the whole space. A natural question is whether an arbitary
ensembles of LOCC-distinguishable orthogonal states satisfies the condition.
We first show that, in this paper, the answer is positive. Then we
generalize it into multipartite systems, and show that \textit{a necessary
condition for LOCC-distinguishability of multipartite orthogonal quantum
states is that the sum of the least numbers of the product states (For
bipartite system, the least number of product states is Schmidit number) of
the orthogonal states is not bigger than the dimensions of the Hilbert space
of the multipartite system}. This necessary condition is very simple and
general, and one can get many cases of indistinguishability by it. It means
that the least number of the product states acts an important role in
distinguishablity of states, and implies that the least number of the
product states may be an good manifestion of quantum nonlocality in some
sense. In fact, entanglement emphases the "amount" of nonlocality, but the
least number of the product states emphases the types of nonlocality. For
example, the known W states and GHZ states have different least number of
the product states, and are different in type.
\end{abstract}

\maketitle

Taking bipartite systems as examples, distinguishing locally orthogonal
quantum states can be discribed as: Alice and Bob hold a part of a quantum
system, which occupies one of $m$ possible orthogonal states $\left\{
\left\vert \Psi _{i}\right\rangle ,i=1,...,m\right\} $. Alice and Bob know
the precise form of these states, but don't know which of these possible
states they actually hold. To distinguish these possible states they will
perform some local operations and classical communication (LOCC): Alice (or
Bob) first measures her part. Then she tells the Bob her measurement result,
according to which Bob measures his part, and so on. With these measurement
results they can exclude some or all possibilities of the system \cite{1}.
Obviously, the possible states can be distinguished if the global
measurements are allowed. But they may not be distinguishable if only LOCCs
are allowed. The fact that some orthogonal quantum states cannot be
distinguished by LOCC is one of the interesting manifesties of non-locality
in quantum mechanics. On the other hand, from the point of view of
informaton, distinguishing locally orthogonal quantum states can also be
imagined as: a information resource Charles owns two particles and encodes
information using $m$ possible orthogonal states of two particles, then
Charles sends one of the particles to Alice and the other to Bob. Alice and
Bob do rounds of LOCC to gain the encoded information. So distinguishing
locally orthogonal quantum states is to gain information by LOCC, in essence.

There are much attentions on distinguishing locally quantum states \cite%
{1,2,22,3,4,watro,5,6,7,v,m} and gainning locally information \cite{pi,ho}.
Bennett et al showed that there are nine orthogonal product states in a $%
3\otimes 3$ system which cannot be distinguished by LOCC \cite{2}. Walgate
et al showed that any two multipartite orthogonal states can be
distinguished by LOCC \cite{1}. For two-qubit systems (or $2\otimes 2$
systems), any three of the four Bell states cannot be distinguished by LOCC
if only a single copy is provided \cite{3}. References \cite{pi,ho} discuss
the rules of gaining locally information. LOCC-distinguishability of
orthogonal states is related to many important directions in quantum
information, such as distillable entanglement \cite{v,m,hsc}, quantum
cloning, hiding information \cite{hady}, quantum channel (See Ref. \cite%
{watro} and references therein), and quantum information basic theory (See
Refs. \cite{mh} and references therein).

In spite many results, the complex nature of LOCC-distinguishability is far
from clear. Now, the known ensembles of orthogonal states which are
LOCC-distinguishable have a common feature that the sum of Schmidit numbers
of the orthogonal states is not bigger than the dimensions of the whole
space. For example, the ensembles of Bell states in Ref. \cite{6} and the
ensembles of orthogonal states in an $2\otimes n$ system \cite{5} have this
feature. A natural question is: whether an arbitary ensemble of
LOCC-distinguishable states has this feature? In this paper, we will show
the answer is positive. We first show that if orthogonal states of arbitrary
ensemble are LOCC-distinguishable, each orthogonal state is a linear
superposition of a set of linearly independent product states, and the
number of the linearly independent product states is not less than the sum
of Schmidit numbers (or the least number of the product states for
multipartite) of the orthogonal states. Then we present a very simple but
powerful ceriterion to judge indistinguishability of states: if the sum of
Schmidit number (or the least number of the product states) of the
orthogonal states is bigger than the dimensions of the space, the states are
LOCC-indistinguishable. With this criterion one can get many cases of
indistinguishability \cite{3}. Finally we discussion the effect of average
entanglement and output entanglement on the indistinguishability of states.
The conclusions may be useful in discussing the distinguishability of
orthogonal quantum states further, understanding the essence of nonlocality
and discussing the distillation of entanglement and hiding information.

Any protocol to distinguish the $m$ possible orthogonal states can be
conceived as many successive rounds of POVMs and communication by Alice and
Bob. The effect of these measurements and communication can be describled by
a set of operators $\{A_{i}\otimes B_{i},i=1,\cdots ,M\}$ acting on the
Alice and Bob's Hilbert space \cite{nilsen}. Each operator $A_{i}($ or $%
B_{i} $ ) is a product of\ positive operators and unitary maps corresponding
to Alice's (or Bob's) measurements and ratations. If an outcome $i$ occures,
the measured state $\left\vert \Psi _{j}\right\rangle $ becomes

\begin{equation}
\left\vert \Psi _{j}\right\rangle \rightarrow A_{i}\otimes B_{i}\left\vert
\Psi _{j}\right\rangle ,
\end{equation}%
where $A_{i}\otimes B_{i}$ satisfies the complete relation

\begin{equation}
\sum_{i}A_{i}^{+}\otimes B_{i}^{+}A_{i}\otimes B_{i}=I.  \label{1}
\end{equation}%
According to the polar and singular value decompositions of operators \cite%
{nilsen, lin}, operators $A_{i},B_{i}$ can be expressed as the product of a
positive operator and an unitary operator, namely 
\begin{equation}
A_{i}=u_{iA}A_{i}^{\prime };\qquad B_{i}=u_{iB}B_{i}^{\prime }  \label{3}
\end{equation}%
\begin{eqnarray}
A_{i}^{\prime } &=&c_{1}^{i}\left\vert \varphi _{1}^{i}\right\rangle
_{A}\left\langle \varphi _{1}^{i}\right\vert +\cdots +c_{N_{a}^{\prime
}}^{i}\left\vert \varphi _{N_{a}^{\prime }}^{i}\right\rangle
_{A}\left\langle \varphi _{N_{a}^{\prime }}^{i}\right\vert  \label{a} \\
0 &<&c_{j}^{i}\leq 1,j=1,\cdots ,N_{a}^{\prime };N_{a}^{\prime }\leq N_{a} 
\nonumber
\end{eqnarray}%
\begin{eqnarray}
B_{i}^{\prime } &=&d_{1}^{i}\left\vert \eta _{1}^{i}\right\rangle
_{B}\left\langle \eta _{1}^{i}\right\vert +\cdots +d_{N_{b}^{\prime
}}^{i}\left\vert \eta _{N_{b}^{\prime }}^{i}\right\rangle _{B}\left\langle
\eta _{N_{b}^{\prime }}^{i}\right\vert  \label{b} \\
0 &<&d_{j}^{i}\leq 1,j=1,\cdots ,N_{b}^{\prime };N_{b}^{\prime }\leq N_{b} 
\nonumber
\end{eqnarray}%
where $u_{iA}$ and $u_{iB}$ are unitary; $A_{i}^{\prime }$ is diagonal
positive operators and filtrations which change the relative weights of
components $\left\vert \varphi _{1}^{i}\right\rangle ,\cdots ,\left\vert
\varphi _{N_{a}}^{i}\right\rangle ,$ and similarly for$\ B_{i}^{\prime }$. $%
N_{a}$ and $N_{b}$ is the dimensions of Alice's and Bob's Hilbert space $%
H_{a},H_{b}$, respectively.

If operators $\{A_{i}\otimes B_{i},i=1,\cdots ,M\}$ can distinguish
perfectively the states $\left\{ \left\vert \Psi _{i}\right\rangle
,i=1,...,m\right\} ,$ then each operator $A_{i}\otimes B_{i}$ corresponding
to outcome $i$ \textquotedblleft indicate\textquotedblright\ only a state,
namely%
\begin{equation}
A_{i(s)}\otimes B_{i(s)}\left\vert \Psi _{i}\right\rangle \neq 0;  \label{d1}
\end{equation}%
\begin{equation}
A_{i(s)}\otimes B_{i(s)}\left\vert \Psi _{j}\right\rangle =0,j\neq i
\label{d2}
\end{equation}%
Eq (\ref{d1}) means $A_{i}\otimes B_{i}$ can indicate $\left\vert \Psi
_{i}\right\rangle ,$ and Eq (\ref{d2}) means $A_{i}\otimes B_{i}$ indicates
only $\left\vert \Psi _{i}\right\rangle .$ Of course, it is possible that a
possible states $\left\vert \Psi _{i}\right\rangle $ may be indicated by
many operators. The lower index $(s)$ of $A_{i}\otimes B_{i}$ denotes many
operators indicating the state $\left\vert \Psi _{i}\right\rangle .$

Furthermore, operator $A_{i}\otimes B_{i}$ indicate only $\left\vert \Psi
_{i}\right\rangle $ means that all states $\left\vert \Psi _{j}\right\rangle
(j\neq i)$ are orthogonal to the subspace spanned by the bases $\left\{
\left\vert \varphi _{l}^{i}\right\rangle _{A}\left\vert \eta
_{k}^{i}\right\rangle _{B},l=1,\cdots ,N_{a}^{\prime };k=1,\cdots
,N_{b}^{\prime }\right\} .$ So if operator $A_{i}\otimes B_{i}$ indicates
only $\left\vert \Psi _{i}\right\rangle ,$ all the following $N_{a}^{\prime
}N_{b}^{\prime }$ one-rank operators $\left\{ c_{l}^{i}\left\vert \varphi
_{l}^{i}\right\rangle _{A}\left\langle \varphi _{l}^{i}\right\vert \otimes
d_{k}^{i}\left\vert \eta _{k}^{i}\right\rangle _{B}\left\langle \eta
_{k}^{i}\right\vert ,l=1,\cdots ,N_{a}^{\prime };k=1,\cdots ,N_{b}^{\prime
}\right\} $ indicate only $\left\vert \Psi _{i}\right\rangle $ (It is
possibel that some of the one-rank operators indicate none of the possible
states). Similarly, the other operators $A_{j}\otimes B_{j}$ $(j\neq i)$
also correspond to similar one-rank operators. Let $\{a_{i}\otimes
b_{i}=e_{i}\left\vert \varphi _{i}\right\rangle _{A}\left\langle \varphi
_{i}\right\vert \otimes \left\vert \eta _{i}\right\rangle _{B}\left\langle
\eta _{i}\right\vert \}\quad (e_{i}>0,i=1,\cdots ,M^{\prime },M^{\prime
}\geq N_{a}N_{b}$ $)$ denote all these one-rank operators. Obviously these
one-rank operators satisfy the complete relation $\sum_{i}a_{i}^{+}\otimes
b_{i}^{+}a_{i}\otimes b_{i}=I_{N_{a}\otimes N_{b}}.$ Moreover, one can
carries out these one-rank operators by doing the consequent projective
measurements unitary ratations after one has carried out the operators $%
\{A_{i}\otimes B_{i},i=1,\cdots ,M\}.$ For example, after one gets output $i$
corresponding to operator $A_{i}\otimes B_{i},$ one can carries out one-rank
operators $\left\{ c_{l}^{i}\left\vert \varphi _{l}^{i}\right\rangle
_{A}\left\langle \varphi _{l}^{i}\right\vert \otimes d_{k}^{i}\left\vert
\eta _{k}^{i}\right\rangle _{B}\left\langle \eta _{k}^{i}\right\vert
,l=1,\cdots ,N_{a}^{\prime };k=1,\cdots ,N_{b}^{\prime }\right\} $ by doing
two unitary ratations $u_{iA}$ and $u_{iB}$, and doing local peojective
measurements$\left\{ \left\vert \varphi _{l}^{i}\right\rangle
_{A}\left\langle \varphi _{l}^{i}\right\vert ,l=1,\cdots ,N_{a}^{\prime
}\right\} $ and $\left\{ \left\vert \eta _{k}^{i}\right\rangle
_{B}\left\langle \eta _{k}^{i}\right\vert ,k=1,\cdots ,N_{b}^{\prime
}\right\} .$ Thus we can have the following Lemma, as shown in Ref. \cite%
{watro}

Lemma: If states $\left\{ \left\vert \Psi _{i}\right\rangle
,i=1,...,m\right\} $ can be distinguished perfectively by $\{A_{i}\otimes
B_{i},i=1,\cdots ,M\}$, they can also be distinguished by a set of one-rank
operators $\{a_{i}\otimes b_{i}\}$.

Since $\sum_{i}a_{i}^{+}\otimes b_{i}^{+}a_{i}\otimes
b_{i}=\sum_{i}e_{i}^{2}\left\vert \varphi _{i}\right\rangle _{A}\left\langle
\varphi _{i}\right\vert \otimes \left\vert \eta _{i}\right\rangle
_{B}\left\langle \eta _{i}\right\vert =I_{N_{a}\otimes N_{b}},$ any state $%
\left\vert \Psi \right\rangle $ in the Hilbert space $H_{a}\otimes H_{b}$ is
the linear superposition of the product states $\{\left\vert \varphi
_{i}\right\rangle _{A}\left\vert \eta _{i}\right\rangle _{B},i=1,\cdots
,M^{\prime }\}$, namely, $\left\vert \Psi \right\rangle
=\sum_{i}e_{i}^{2}\left\langle \varphi _{i}\right\vert _{A}\left\langle \eta
_{i}\right\vert _{B}\Psi \rangle $ $\left\vert \varphi _{i}\right\rangle
_{A}\left\vert \eta _{i}\right\rangle _{B}.$ Owing to a operator $%
a_{i}\otimes b_{i}$ indicates no more than a possible state, let's name the
product states $\{\left\vert \varphi _{i}\right\rangle _{A}\left\vert \eta
_{i}\right\rangle _{B},i=1,\cdots ,M^{\prime }\}$ indicating product states
(IPS). Each IPS are orthogonal to all possible states except for no more
than one possible state.

Since non-orthogonal measurements are allowed, not all IPS are linearly
indepentent to each other, in general. But we can always find $N_{a}N_{b}$
linearly indepedent IPS (LIIPS) $\left\vert LIIPS\right\rangle
_{j}(j=1,\cdots ,N_{a}N_{b})$ such that they form a set of complete
nonorthogonal product bases of the space $H_{a}\otimes H_{b}$. All states in
the space $H_{a}\otimes H_{b}$ is the linear superposition of the LIIPS.
Obviously, the number of LIIPS in a state $\left\vert \Psi _{i}\right\rangle 
$ is at least the Schmidt number of the state. Furthermore, if states $%
\left\{ \left\vert \Psi _{i}\right\rangle ,i=1,...,m\right\} $ can be
distinguished perfectively by LOCC, each LIIPS\ exists in no more than one
possible state. So if states $\left\{ \left\vert \Psi _{i}\right\rangle
,i=1,...,m\right\} $ can be distinguished perfectively by LOCC, the sum of
the numbers of the LIIPS in all possible state is not less than the sum of
the Schmidt numbers of the possible states. Thus we have proven following
Theorem.

Theorem 1: If states $\left\{ \left\vert \Psi _{i}\right\rangle
,i=1,...,m\right\} $ can be distinguished perfectively by LOCC, the number
of LIIPS in a possible state $\left\vert \Psi _{i}\right\rangle $ is not
less than the Schmidt number of the $\left\vert \Psi _{i}\right\rangle ,$
and then the number of LIIPS in all possible state is not less than the sum
of the Schmidt numbers of the possible states.

Frome Theorem 1 we can get a interesting conclusion:

Theorem 2: If a set of orthogonal states $\left\{ \left\vert \Psi
_{i}\right\rangle ,i=1,...,m\right\} $ in a Hilbert space shared by Alice
and Bob can be distinguished perfectively by LOCC, the sum of Schmidt
numbers of the states is not bigger than the dimensions of the space.

Proof: The proof is very simple. If the orthogonal states $\left\{
\left\vert \Psi _{i}\right\rangle ,i=1,...,m\right\} $ are LOCC
distinguishable, and the sum of Schmidt numbers of the states is bigger than
the dimensions of the space, then from Theorem 1 we can follow that the
number of LIIPS is bigger than the dimensions $N_{a}N_{b}$ of the whole
space $H_{a}\otimes H_{b}.$ This is impossible, and then completes the proof.

In the discussion above, the Schmidt number of a bipartite pure state is, in
essence, the least number of product states of the pure state. So the
results and their proof of Theorem 1 and 2 can be generalized into
multi-partite system, obviously, if we replace "Schmidt number" by "the
least number of product states" in a possible state. Thus we have that: 
\textit{a necessary condition for distinguishability of multipartite
orthogonal states is that the sum of the least numbers of product states of
the orthogonal states is not bigger than the dimensions of Hilbert space of
the multipartite system.}

From the theorem 2 one can get the many interesting cases. For example, for $%
n\otimes n$ systems one cannot distinguish deterministically $n+1$ states,
each of which has Schmidt number $n$ \cite{nath}; for $n\otimes n$ systems,
if one can distinguish $n^{2}$ orthogonal states, these states must be
orthogonal product vectors as shown in Ref. \cite{7}; for three qubits
systems, three W-type orthogonal states are LOCC-indistinguishable (W-type
states have the form of $a\left\vert 001\right\rangle +b\left\vert
010\right\rangle +c\left\vert 100\right\rangle )$.

As stated in the beginning, distinguishing locally orthogonal quantum states
is related to gainning information by LOCC. Charles encodes information
using orthogonal states $\left\{ \left\vert \Psi _{i}\right\rangle ,\text{ }%
p_{i},i=1,...,m,\right\} $ of two particles A and B, where $p_{i}$ is the
probability $\left\vert \Psi _{i}\right\rangle $ occures. Then Charles sends
one of the particles to Alice and the other to Bob. Alice and Bob try to
gain the encoded information by LOCC. The locally accessible information
from the ensemble $\sigma =\left\{ \left\vert \Psi _{i}\right\rangle ,\text{ 
}p_{i},i=1,...,m,\right\} $ is limited by \cite{pi}

\begin{equation}
I_{acc}^{LOCC}(\sigma )\leq \ln (N_{a}N_{b})-E,  \label{i}
\end{equation}%
and further by \cite{ho}

\begin{equation}
I_{acc}^{LOCC}(\sigma )\leq \ln (N_{a}N_{b})-E-E_{f},  \label{i1}
\end{equation}%
where $E$ is the average of the entanglement of $\sigma ;$ $E_{f}$ is the
average entanglement of the output.

Now we give a qualitative explanation of Eq (\ref{i}) and (\ref{i1}). The
state of the infinite copies of the ensemble, $\sigma ^{\otimes
n}(n\rightarrow \infty ),$ is a mixture of $2^{nS(\sigma )}$ "likely"
pure-states-strings with equal probability \cite{ben}, where $S(\sigma )$ is
von Neumann entropy. All Schmidt coefficient of the pure-states-strings are
equal, and the Schmidt number of a string are $2^{nE},$ where $nE$ is the
entanglement of each string. We now encode locally accessible information
using these strings, namely, encode a locally accessible singnal using a
string. From Theorem 1 and 2, one needs at least $2^{nE}$ LIIPS to encode a
locally accessible singnal. So one can encode $\frac{(N_{a}N_{b})^{\otimes n}%
}{2^{nE}}$ singnal at most in the spapce $(H_{a}\otimes H_{b})^{\otimes n}$.
Namely $I_{acc}^{LOCC}(\sigma ^{\otimes n})\leq \ln \frac{%
(N_{a}N_{b})^{\otimes n}}{2^{nE}}=n(\ln (N_{a}N_{b})-E).$ Obviously, $%
nI_{acc}^{LOCC}(\sigma )\leq I_{acc}^{LOCC}(\sigma ^{\otimes n}),$ thus we
get Eq (\ref{i}).

For one-rank operators, each output is a product state. If the output states
are not product states, but entanglement states with average entanglement
amount $E_{f},$ we can also explain Eq (\ref{i1}) by considering $\sigma
^{\otimes n}(n\rightarrow \infty ).$ After one has finished the operators to
distinguish strings, one can get pure states the entanglement of each of
which is $nE_{f}.$ In this case the operators $\{A_{i}\otimes
B_{i},i=1,\cdots ,M\}$, which can distinguish the strings, are not the
one-rank operators, but at least $2^{2nE_{f}}$-rank ones. $A_{i}\otimes
B_{i} $ project out an $2^{nE_{f}}\otimes 2^{nE_{f}}$ dimensions space, $%
2^{2nE_{f}}$ bases of which is LIIPS. Moreover, each string is indicated by
at least $2^{nE-nE_{f}}$ operators of $\{A_{i}\otimes B_{i},i=1,\cdots ,M\}$%
. So the LIIPS of each string is at least $%
2^{nE-nE_{f}}2^{2nE_{f}}=2^{nE+nE_{f}},$ and then Alice and Bob can encode
at most $\ln \frac{(N_{a}N_{b})^{\otimes n}}{2^{nE+nE_{f}}}=n(\ln
(N_{a}N_{b})-E-E_{f})$ bits locally accessible information in the space $%
(H_{a}\otimes H_{b})^{\otimes n}$. Obviously, $nI_{acc}^{LOCC}(\sigma )\leq
I_{acc}^{LOCC}(\sigma ^{\otimes n}),$ thus we get Eq (\ref{i1}).

LOCC-indistinguishability seem be related to entanglement. For example, Eq (%
\ref{i}) and (\ref{i1}) imply that entanglement results in the
indistinguishability. On the other hand, there are ensembles of
LOCC-indistinguishable orthogonal \textit{product} states \cite{2,22}.
Moreover, one can destroy LOCC-distinguishability by reducing the average
entanglement of the ensemble of states \cite{7}. The two "opposite" results
can be expained as shown in this paper: LOCC-indistinguishability is not
related directly to average entanglement of orthogonal states, but to the
least number of product states of the orthogonal states.
LOCC-indistinguishability is related directly to average entanglement of
orthogonal states at the limit of infinite number of copies of the ensemble.

In summary, we present a necessary condition for distinguishability of
multipartite orthogonal quantum states, which is simple and general. With
this condition one can get many cases of indistinguishability. These results
mean that the least number of the product states acts an important role in
distinguishablity of states. This implies that the least number of the
product states may be an good manifestion of quantum nonlocality in some
sense. In fact, entanglement emphases the "amount" of nonlocality, but the
least number of the product states emphases the types of nonlocality. For
example, the known W states and GHZ states have different least number of
the product states, and are different in type.

The results in this paper open three interesting questions: 1. For ensembles
of states which satisfy the condition in Theorem 2 but are still
indistinguishable, whether the indistinguishability of the states results
from the existence of the subspace the projections of the states on the
subspace do not satisfy the condition? More precisely, is the following
result ture? --- if there is a subspace the sum of the Schmidt number of the
projections of the states on the subspace is bigger than the dimensions of
the subspace, the states are LOCC-indistinguishable. This supposed result is
stronger than Theorem 2, obviously. If the supposition is ture, it
approaches to a necessary and sufficient condition of
LOCC-distinguishaility. 2. To our knowledge, no ensembles of bipartite
states which can be distinguished by POVMs but not by projective measurement
has been found, while the ensemble of multipartite states has been found 
\cite{chen}. Is it ture that if bipartite states can be distinguished by
POVMs, they can be distinguished by projective measurements? 3. Eqs. (\ref{i}%
) and (\ref{i1}) can be generalized into multipartite if we re-define $E$
and $E_{f}$ corresponding to the least number of product states in a pure
state, in principle. But the detailed investigations into the generalization
are necessary.

This work is supported by the National Natural Science foundation (No.
Grants 10404039, 10204020), the Chinese National Fundamental Research
Program (2001CB309300), the Innovation funds from Chinese of Academy of
Sciences, and the China Postdoctoral Science Foundation.


\begin{references}
\bibitem{1}  J.Walgate, A.J.Short, L.Hardy and V.Vedral,Phys.Rev.Lett.85,4972 (2000)

\bibitem{2} C. H. Bennett, D.P. DiVincenzo, C.A. Fuchs, T.Mor, E.Rains,P.W. Shor, J.A. Smolin, and W.K. Wootters, Phys. Rev. A 59,1070 (1999) or
quant-ph/9804053.

\bibitem{22} C. H. Bennett, D.P. DiVincenzo, T.Mor, P.W. Shor, J.A. Smolin, and B. W. Terhal, Phys. Rev. Lett 82,5385 (1999).

\bibitem{3} M. Horodecki, P. Horodecki, and R. Horodecki, Acta Physica Slovaca, 48, (1998) 141, or quant-ph/9805072

\bibitem{4} S. Virmani, M.F. Sacchi, M.B. Plenio and D. Markham, Physics Letters A 288, 62-68 (2001);Y.-X.Chen and D.Yang, Phys.Rev.A 64, 064303 (2001)

\bibitem{watro}  John Watrous, Phys. Rev. Lett 95, 080505 (2005)

\bibitem{5} J. Walgate and L. Hardy, Phys. Rev. Lett 89, 147901 (2002)

\bibitem{6} S.Ghosh, G.Kar, A.Roy, A.Sen and U.Sen, Phys.Rev.Lett.87, 277902 (2001); S. Ghosh, G.Kar, A.Roy, D.Sarkar, A.Sen(De) and U.Sen, Phys. Rev. A 65, 062307

\bibitem{7} M. Horodecki, A. Sen (De), U. Sen and K. Horodecki, Phys.Rev.Lett.90, 047902 (2003)

\bibitem{v}  V. Vedral and M. B. Plenio, Phys. Rew A 57, 1619 (1998);

\bibitem{m}  V. Vedral M. B. Plenio, K. Jocobs and P. L. Knight, Phys. Rew A
56, 4452 (1997)

\bibitem{pi}  P. Badziag, M. Horodecki, A. Sen(De) and U. Sen, Phys. Rev.
Lett. 91, 117901 (2003).

\bibitem{ho}  M. Horodecki, J. Oppenheim, A. Sen(De) and U. Sen, Phys. Rev.Lett. 93, 170503 (2004).

\bibitem{hsc}  S. Ghosh, P. Joag, G. Kar, S. Kunkri and A. Roy,arXiv:quant-ph/0403134 (2004); P.- X Chen and C.-Z Li, Quant. Inf and
Comp,V3 203 (2003).

\bibitem{hady}  P. Hayden, D. Leung, and G. Smith Phys. Rev. A 71, 062339 (2005).

\bibitem{mh}  M. Horodecki, J. Oppenheim, A. Winter Nature 436, 673 (2005); M. Horodecki et al, Phys. Rev. A 71, 062307 (2005)

\bibitem{nilsen}  M. A. Nielsen and I. L. Chuang, Quantum Computation and Quantum Information (Cambridge University Press, Cambridge),pp. 78-79; 

\bibitem{lin} M. Horodecki, P. Horodecki, and R. Horodecki,Phys. Rev. Lett. 80, 5239 (1998);N. Linden, S. Massar and S. Popescu, Phys. Rev. Lett. 81, 3279 (1998);W. K. Wootters, Phys. Rev. Lett. 80,2245 (1998)

\bibitem{nath}  M. Nathanson, J. Math. Phys. (N.Y.) 46, 062103 (2005).

\bibitem{ben}  C. H. Bennett, G. Brassard, S. Popescu and B. Schumacher, Phys. Rev. A 53, 2046 (1996).

\bibitem{chen}  P.-X.Chen and C.-Z Li, Phys.Rev.A 70, 022306(2004).

\end{references}
\end{document}